\documentstyle[12pt]{article}

\newcommand{\be}{\begin{equation}}
\newcommand{\ee}{\end{equation}}
\newcommand{\bea}{\begin{eqnarray}}
\newcommand{\eea}{\end{eqnarray}}

\newcommand{\p}[1]{(\ref{#1})}

 at 14.4truept
 at 14.4truept

\def\Z{{\mathbf  Z}}
\def\C{{\mathbf  C}}
\def\pa{\partial}
\def\ker{{{\mathrm Ker}({\mathrm  ad\,} \Lambda)}}
\def\im{{{\mathrm  Im}({\mathrm  ad\,} \Lambda) }}
\def\A{{\cal A}}
\def\G{{\cal G}}
\def\N{{\cal N}}
\def\L{{\cal L}}
\def\H{{\cal H}}
\def\P{{\cal P}}
\def\R{{\cal R}}
\def\D{{\cal D}}

\def\K{{\cal K}}
\def\M{{\cal M}}
\begin{document}

\thispagestyle{empty}

\setcounter{page}{0}
\renewcommand{\thefootnote}{\fnsymbol{footnote}}
\fnsymbol{footnote}
\rightline{ENSLAPP-L-658/97\break}
\rightline{DIAS-STP-97-12\break}

\vspace{.4in}

\begin{center} \Large \bf Nonstandard Drinfeld-Sokolov reduction\\

\end{center}

\vspace{.1in}

\begin{center}
F.~Delduc${}^{(a)}$, 
L.~Feh\'er${}^{(b)}$\protect\footnote{
Permanent  address: 
Department of  Theoretical Physics,  
J\'ozsef Attila University, H-6720
Szeged, Hungary.}
and L.~Gallot${}^{(a)}$\\

\vspace{0.2in}

${}^{(a)}${\em Laboratoire de Physique Th\'eorique\footnote{\rm  ENSLAPP, 
URA 14-36 du
CNRS, associ\'e \`a l'E.N.S. de Lyon et au L.A.P.P.}\\
ENS Lyon,  46 all\'ee d'Italie\\
F-69364 Lyon Cedex 07, France}

\vspace{0.2in}

${}^{(b)}${\em Dublin Institute for Advanced Studies\\
10 Burlington Road\\
Dublin 4, Ireland}
\end{center}

\vspace{.2in}

{\parindent=25pt
\narrower\smallskip\noindent

\small
\noindent
{\bf  Abstract.}\quad
Subject to some conditions,
the input data for the Drinfeld-Sokolov construction of KdV type 
hierarchies is a quadruplet $(\A,\Lambda, d_1, d_0)$, 
where the $d_i$ are $\Z$-gradations of a loop   algebra $\A$ 
and $\Lambda\in \A$ is a semisimple element of nonzero $d_1$-grade.
A new  sufficient condition on the quadruplet 
under which the construction works is proposed and examples are presented.
The proposal relies on 
splitting the $d_1$-grade zero part of  $\A$ into a
vector space  direct sum of two subalgebras. 
This permits one to interpret certain 
Gelfand-Dickey type systems associated with a nonstandard splitting 
of the algebra of pseudo-differential operators in the Drinfeld-Sokolov 
framework. 
}

\newpage

\section{Introduction}
\setcounter{equation}{0}

\renewcommand{\thefootnote}{\arabic{footnote}}
\setcounter{footnote}{0}

Developing the ideas of the pioneering papers \cite{DS,Wi},
recently a general Lie algebraic 
framework has been established in which to construct
generalized KdV and modified KdV type integrable hierarchies 
\cite{McI,Prin1,Prin2}.
This formalism contains many interesting 
systems as special cases \cite{DF,FM}.
However, there  exist some well-known  
systems which do not seem to fit in  the approach as has been developed 
so far. 
For example, while the standard Gelfand-Dickey hierarchies
\cite{Di}
\be
{\pa \over \pa t_m} L = 
\left[ \left(L^{m\over n}\right)_{\geq 0}, L\right]  
\quad\hbox{for}\quad
L= \pa^n +\sum_{i=1}^n u_i \pa^{n-i} 
\label{1.1}\ee
have a well-known interpretation \cite{DS}, which motivated the whole theory,
their  `nonstandard'  counterparts \cite{kuper,OS,Oevel2} defined by 
\be
{\pa \over \pa t_m} L= \left[ \left(L^{m\over n-1}\right)_{\geq 1}, L\right]  
\quad\hbox{for}\quad
L= \pa^{n-1} +\sum_{i=1}^{n-1} u_i \pa^{n-1-i}+ \pa^{-1} u_{n} 
\label{1.2}\ee
so far resisted a similar interpretation.

In this paper we propose an   
extension of the above mentioned Lie algebraic framework of constructing 
integrable hierarchies.
This  will prove general enough to contain
the nonstandard Gelfand-Dickey hierarchies as special cases.

The Drinfeld-Sokolov construction relies on the use of a classical 
r-matrix \cite{STS} given by the difference of two projectors,
$\R= {1\over 2}(\P_\alpha - \P_\beta)$, 
associated with a splitting $\A=\alpha +\beta$ of an affine 
Lie algebra $\A$. 
So far it has been  assumed \cite{DS,Wi,McI,Prin1,Prin2}
that the subalgebras $\alpha, \beta \subset \A$
have the form 
$\alpha = \A^{\geq 0}$ and 
$\beta = \A^{<0}$
in terms of a $\Z$-gradation $\A=\oplus_{n\in \Z} \A^n$ of $\A$. 
The essence of the proposal of this paper will be to use a more
general r-matrix obtained by further splitting $\A^0$.
That is we shall use 
\be
\R= {1\over 2}(\P_\alpha - \P_\beta)
\quad\hbox{where}\quad
\alpha = \A^{>0} + \alpha^0,
\quad
\beta = \A^{<0} + \beta^0 
\label{1.3}\ee
in correspondence with a splitting
$\A^0 = \alpha^0 + \beta^0$
subject to certain conditions.
The standard  construction 
will be recovered for $\beta^0=\{ 0\}$. 

The standard and nonstandard Gelfand-Dickey 
systems correspond to 
two splittings of the algebra of pseudo-differential
operators that differ in the scalars being added to the
subalgebras of purely differential or purely integral 
operators,
which is reminiscent of the manipulation with the splittings 
in our construction.
Since  the nonstandard Gelfand-Dickey systems are recovered from it,
we sometimes refer to our  construction as the 
`nonstandard Drinfeld-Sokolov construction'.
However, it should be stressed that 
our construction is  essentially just 
the standard one implemented  
under weakened conditions on the input data. 

The layout of the  paper is as follows.
Section 2 is devoted to explaining the nonstandard 
Drinfeld-Sokolov construction. In subsection 2.1 
modified KdV type systems are dealt with.
The construction of KdV type systems is described in 
subsection 2.2.
The nonstandard Gelfand-Dickey systems are derived
as examples in section 3.
Section 4 contains  our conclusions.
Throughout the paper, 
proofs are often omitted or 
kept short since they are similar to those in the standard case.

\section{A general construction of integrable hierarchies}
\setcounter{equation}{0}

The standard construction \cite{DS,Wi,McI,Prin1,Prin2} 
associates a modified KdV type
system with a triplet $(\A,  \Lambda, d_1  )$ and a KdV type system
with a quadruplet $(\A,\Lambda, d_1, d_0)$, where the 
$d_i$ ($i=0,1$)  are $\Z$-gradations of an affine Lie algebra $\A$,
$\Lambda\in \A$ 
is a semisimple element of $d_1$-grade $k>0$,  and some further
conditions hold in the KdV case.
Below we present a generalization of the standard construction
based on a splitting of the $d_1$-grade zero subalgebra
$\A^0$ into a direct sum of two subalgebras of a certain form,
and weakened conditions on the gradations $d_i$.

\subsection{Modified KdV type systems}

Let  $\A$  be an affine Lie algebra with vanishing center, that is  
a twisted loop algebra 
\be
\A=\ell(\G,\tau)\subset \G\otimes \C [\lambda,\lambda^{-1}]
\label{2.1}\ee
attached to a finite dimensional complex simple Lie algebra $\G$ 
with an automorphism $\tau$ of finite order \cite{Kac}.
Let $\A=\oplus_{n\in \Z} \A^n$ denote a  $\Z$-gradation  of $\A$
given by the eigensubspaces of a derivation $d_1$ of $\A$ as  
$d_1(X)=n X$ for $X\in \A^n$. 
Consider a semisimple element $\Lambda \in \A^k$ for some $k>0$,
and two subalgebras $\alpha^0$, $\beta^0$ 
of $\A^0$ in such a way that 
\be
\beta^0 \subset \ker
\quad\hbox{and}\quad 
\A^0 = \alpha^0 + \beta^0 
\label{2.2}\ee
is a linear direct sum decomposition. 
The subsequent construction, 
which reduces to that in \cite{DS,Wi,McI,Prin1,Prin2} for $\beta^0=\{ 0\}$, 
defines a mod-KdV type system for any choice of 
the data $(\A,  \Lambda, d_1; \alpha^0, \beta^0)$.
By definition, the phase space of this 
system is the manifold $\Theta$
of first order differential operators given by
\be
\Theta:=\left\{ \L=\pa_x + \theta(x) +\Lambda\,\vert\,
\theta(x)\in \A^{<k} \cap \A^{\geq 0}\,\right\}.
\label{2.3}\ee
We use the notation $\A^{<k}=\oplus_{n<k} \A^n$ etc.
Since $\A^{<k}\cap \A^{\geq 0}$ is a finite dimensional space,
the field $\theta(x)$ encompasses a finite
number of complex valued fields depending on the one-dimensional
space variable $x$.
We wish  to exhibit a family of compatible evolution equations on $\Theta$
labelled by  the  graded basis elements of 
\be
{\cal C}(\Lambda):=\left(\mbox{\rm Cent}\, \ker\right)^{\geq 0},
\label{2.4}\ee
which is the positively graded part of the center of the Lie
algebra $\ker$.
For this we shall use the well-known `formal dressing procedure' 
based on the linear direct sum decomposition
\be
\A=\ker +\im,
\qquad
\ker\cap\im =\{ 0\},
\label{2.5}\ee
whose existence is guaranteed by the semisimplicity of $\Lambda$.
We next recall the main points of this  procedure in a slightly
more general context than required in this subsection.

\medskip
\noindent{\bf Lemma 1.}
{\em Let $j(x)\in \A^{<k}$ be  an arbitrary  
formal series with smooth component functions.
Consider the equation  
\begin{equation}
\L:= (\pa_x + j(x) + \Lambda)  \mapsto e^{{\mathrm{ad}} F}(\L)=
\pa_x + h(x) + \Lambda,
\label{2.6}\end{equation}
where  $F(x)$ and $h(x)$ are required to be formal series 
\begin{equation}
F(x)\in \A^{<0},
\qquad
h(x)\in \left(\ker\right)^{<k}.
\label{2.7}\end{equation}
Then (\ref{2.6})  can be solved for $F(x)$ and 
in terms of a  particular solution $F_0(x)$  
the general solution is determined by 
\be
e^{{\mathrm{ad}} F}=e^{{\mathrm{ad}} K} e^{{\mathrm{ad}} F_0}, 
\label{2.8}\ee
where $K(x)\in \left(\ker\right)^{<0}$ is arbitrary.
There is a unique solution 
$F(j(x))\in \left(\im\right)^{<0}$, whose components are 
differential polynomials in the components of $j(x)$.}
\medskip

Thanks to the  lemma, which goes back to Drinfeld-Sokolov \cite{DS}
(see also\cite{McI,Prin1}), for any constant $b\in {\cal C}(\Lambda)$
and any function $j(x)\in \A^{<k}$ one  can define
 \begin{equation}
B_b (j):=e^{-{\mathrm{ad}} F(j) } (b).
\label{2.9}\end{equation}
The components of $B_b(j)$  are uniquely determined differential
polynomials in the components of $j$ and one has  
 $[B_b(j), (\pa_x + j+\Lambda)]=0$ as a result of 
$[b, (\pa_x + h +\Lambda)]=0$.
For later purpose,  note also that the formula
\be
\H_b(j):=\int dx\, h_b(j(x)) 
\qquad\hbox{with}\quad h_b(j(x)):=\langle b, h(j(x))\rangle 
\label{2.10}\ee
yields a well-defined functional of $j(x)\in \A^{<k}$ if we assume
that the integral of a total derivative is zero. 
Here $\langle\ ,\ \rangle$ is a nondegenerate,
invariant, symmetric bilinear form on $\A$. 
(Such a bilinear form exists and is unique up to a constant; 
the density $h_b(j(x))$ is well-defined only up to a total
derivative in general.)
According to a standard calculation \cite{DS},
the functional derivative of $\H_b(j)$ defined using
this bilinear form can be taken to be  
\be
{\delta \H_b \over \delta j}=B_b(j) 
\qquad
\hbox{for}\quad j(x)\in \A^{<k},\quad b\in {\cal C}(\Lambda).
\label{2.11}\ee

Since the conditions of Lemma 1 hold on $\Theta$ in (\ref{2.3}),
we can apply the dressing procedure to construct an integrable hierarchy 
on this manifold.
For this  we now introduce the splitting, 
\be
\A = \alpha + \beta \quad\hbox{with}\quad
\alpha = \A^{>0} + \alpha^0,
\quad
\beta=\A^{<0} + \beta^0
\label{2.12}\ee 
using (\ref{2.2}), 
and the corresponding classical r-matrix of $\A$, 
\be
\R={1\over 2}{(\P_\alpha - \P_\beta)},
\label{2.13}\ee
where $\P_\alpha$, $\P_\beta$ project $\A$ 
 onto the respective subalgebras $\alpha$, $\beta$.
By definition, 
the evolution equation associated with 
$b\in {\cal C}(\Lambda)$ is given by
the  following vector field ${\pa \over \pa t_b}$ on $\Theta$:
\be
{\pa   \over \pa t_b }\theta:=\left[ \R(B_b(\theta )), \L\right]
=\left[ \P_\alpha (B_b(\theta )), \L\right]
=-\left[\P_\beta(B_b(\theta)), \L\right]
\quad\hbox{at}\quad \L=(\pa_x + \theta +\Lambda)\in \Theta.
\label{2.14}\ee
The second and third equalities hold since 
$[B_b(\theta), \L]=0$.
It is easy to see from these equalities that 
${\pa  \over \pa t_b } \theta$ is a differential polynomial 
in the components of $\theta(x)$ with values in 
$\A^{<k}\cap\A^{\geq 0}$, 
which guarantees that (\ref{2.14}) makes sense 
as a vector field  on $\Theta$. 
Using that $[B_a(\theta), B_b(\theta)]=0$ for all 
$a,b\in {\cal C}(\Lambda)$
and that as a 
consequence of (\ref{2.14})  
${\pa \over \pa t_a} B_b(\theta)=[\R(B_a(\theta)), B_b(\theta)]$, 
together with the modified classical Yang-Baxter equation \cite{STS}
for $\R$, it can be shown that the vector fields 
associated with different elements of ${\cal C}(\Lambda)$ commute:
\be
\left[ {\pa \over \pa t_a}\,,\, {\pa \over \pa t_{b}}\right]=0
\qquad \forall\,  a,b\in {\cal C}(\Lambda).
\label{2.15}\ee
The functions $h_b(\theta)$ are of course 
the densities of corresponding conserved currents.

Having presented the algebraic construction of the 
hierarchy of evolution equations on $\Theta$, now  
we  describe the  hamiltonian formulation of these equations. 
For this it will be convenient to first introduce a Poisson bracket
on the local functionals on the space
\begin{equation}
\M:=\{ \L=\pa_x + j(x) + \Lambda \,\vert \, j(x)\in \A^{<k}\} 
\label{2.16}\end{equation}
which contains $\Theta$.  
A local functional $f: \M\rightarrow \C$ is given by 
$f(j)=\int dx\, p(x, j(x),\ldots, j^{(n)}(x))$,
where $p$ is a differential polynomial in the components of $j$
whose coefficients are smooth functions of $x$. 
We  let ${\delta f\over \delta j(x)}\in \A$ denote the 
functional derivative of $f$, and  
 introduce  the  r-bracket
\be
[X,Y]_\R=  [\R X, Y] +  [X, \R Y]
\qquad 
 X, Y \in \A.
\label{2.17}\ee
Imposing, for example,  periodic  boundary condition on $j(x)$,
the following formula  defines a Poisson bracket on 
the local functionals: 
\be
\{ f, g\}_\R (j):=\int dx\, 
\left\langle j+\Lambda,
\left[  {\delta f\over \delta j} ,
 {\delta g\over \delta j}\right]_\R\right\rangle 
-\left\langle \R {\delta f\over \delta j},
\pa_x  {\delta g\over \delta j} \right\rangle
- \left\langle  {\delta f\over \delta j},
\pa_x \R {\delta g\over \delta j} \right\rangle .
\label{2.18}\ee
The hamiltonian vector field, $\delta_f$,
corresponding to $f$ is given by 
\be
\delta_f j= [\R {\delta f\over \delta j}, \L]+ \R^t [
{\delta f \over \delta j}, \L]
\qquad \hbox{at}\quad 
 \L=(\pa_x + j +\Lambda)\in \M.
 \label{2.19}\ee
Using the projectors associated with the decomposition
$\A = \alpha^\perp +\beta^\perp$, where $\alpha^\perp$, $\beta^\perp$
denote the annihilators of $\alpha$, $\beta$ in $\A$,
we have $\R^t={1\over 2}(\P_{\beta^\perp} - \P_{\alpha^\perp})$.
More explicitly, 
the hamiltonian vector field reads as 
\be
\delta_f j = 
\P_{\alpha^\perp} [j+\Lambda, \P_\beta {\delta f \over \delta j}]
-\P_{\beta^\perp}[j+\Lambda, \P_\alpha  {\delta f \over \delta j}] 
+ \P_{\alpha^\perp} \P_\beta \pa_x {\delta f \over \delta j}- \P_{\beta^\perp}
\P_\alpha \pa_x {\delta f \over \delta j}.
\label{2.20}\ee

We now have two important statements to make.  
The first is that  $\Theta\subset \M$ is a Poisson submanifold,
and therefore one can trivially restrict the Poisson bracket to $\Theta$. 
The second is that  
the flow in (\ref{2.14}) is hamiltonian
with respect to the Poisson bracket $\{\ ,\ \}_\R$  on $\Theta$ and the
Hamiltonian    $\H_b(\theta)$ obtained by restricting (\ref{2.10}) 
to $\Theta$.
The first statement requires one to check that if 
$j(x)\in \A^{<k}\cap \A^{\geq 0}$
then $\delta_f j(x)$ lies in the same space,
which is readily done with the aid of (\ref{2.20}). 
The second statement follows by combining (\ref{2.11}) with (\ref{2.19}).

\medskip
\noindent
{\em Remark 1.}
One can naturally extend the definition of the commuting vector fields 
${\pa \over \pa t_b}$ to the whole manifold $\M$ in (\ref{2.16}) 
on which Lemma 1 applies. 
The flows of the resulting hierarchy on $\M$ can be written
in hamiltonian form using $\{\ ,\ \}_\R$ in (\ref{2.18}) 
and the Hamiltonian $\H_b(j)$ in (\ref{2.10}).
This system on $\M$ is 
conceptually useful to consider since
both the mod-KdV type and (as we shall see) the KdV type
systems are reductions of it.
In general, it is an interesting problem to find 
all consistent subsystems of the hierarchy  on $\M$
that involve finitely many independent fields.

\subsection{KdV type systems}

We below describe a construction that yields systems
that we 
call
 `systems of KdV type'. 
The construction requires that data of the 
form $(\A, \Lambda, d_1, d_0; \alpha^0, \beta^0)$ be given, where 
$\A$, $\Lambda$, $d_1$ satisfy the previous conditions and
$d_0$ is another $\Z$-gradation of $\A$.
There are further conditions on the data that we specify  next.

We first of all assume that the two $\Z$-gradations of $\A$ 
are compatible, which means that $[d_0, d_1]=0$ and we have 
a bi-gradation  of $\A$:
\be
\A=\oplus_{m,n\in \Z}\, \A_m^n,
\qquad
\A_m^n:=\{\,X\in \A\,\vert\, d_1(X)=nX,
\,\,\, d_0(X)=mX\,\},
\label{2.21}\ee
where superscripts/subscripts denote $d_1/d_0$-grades.
We need the nondegeneracy condition 
\begin{equation}
\ker\cap \A_0^{<0}=\{0\}, 
\label{2.22}\end{equation}
which is a nontrivial condition if $\A_0^{<0}\neq \{0\}$.
We finally suppose  that 
\be
\A^{>0}\subset \A_{\geq 0},
\qquad
\A^{<0}\subset \A_{\leq 0},
\label{2.23}\ee
and  require a splitting of $\A^0$ into a vector space
direct sum of subalgebras  of the form 
\be
\A^0 = \alpha^0 + \beta^0,
\qquad 
\alpha^0 = \A_{\geq 0}^0,
\qquad 
\beta^0 \subset \ker.
\label{2.24}\ee 
The conditions on the two gradations used in \cite{Prin1} 
(see also \cite{McI}) are stronger than 
(\ref{2.23}) in that they include the additional condition 
$\A^0 \subset \A_0$. This extra condition guarantees the existence
of a  splitting of the form  (\ref{2.24}), given by
$\alpha^0 = \A^0$, $\beta^0=\{0\}$.  
In general,  the
existence of a subalgebra $\beta^0 \subset \ker$
which is complementary to $\A^0_{\geq 0}$ in $\A^0$ is a
nontrivial question.
As is easy to see, a necessary condition for the existence 
of such a subalgebra is that
\be
\A^0_{>0} \cap \left(\ker\right)^\perp = \{0\},
\ee
which in particular implies that  
\be 
{\mathrm{dim}}\left(\A^0_{<0} \right)
\leq {\mathrm{dim}}\left(\A^0 \cap \ker\right).
\ee
In the derivation of the system (\ref{1.2}) presented
 in subsection 3.2, we have  
\be
\beta^0= \A^0 \cap \ker.
\ee

Our construction of KdV type systems 
will proceed quite similarly to the
standard one, except that we shall use the r-matrix 
$\R$ given by (2.12), (2.13) together with (2.24) to define 
the commuting vector fields.
In the case when the condition $\A^0 \subset \A_0$ is satisfied 
the construction  reduces to the standard one. 

By definition, the phase space of the KdV type system is the factor space 
$Q/\N$, 
where
\be
Q:=\left\{ \L=\pa_x + q(x) +\Lambda\,\vert\,
q(x)\in \A^{<k}\cap \A_{\geq 0}\,\right\}
\label{2.25}\ee
and $\N$ is the group of `gauge transformations' $e^\gamma$ acting
on $Q$ according to
\be
e^\gamma: \L\mapsto e^{{\mathrm{ad}} \gamma}\left(\L\right)=
e^\gamma \L e^{-\gamma},
\qquad
\L\in Q,
\quad
\gamma(x)\in \A_0^{<0}.
\label{2.26}\ee

Let us  present a  model of $Q/\N$.
Take a $d_1$-graded vector
space  $V$
appearing in a direct sum decomposition
\begin{equation}
\A^{<k} \cap \A_{\geq 0}
=\left[\Lambda, \A_0^{<0}\right] + V.
\label{2.27}\end{equation}
Define   $Q_V\subset Q$  by
\be
Q_V:=\left\{ \L=\pa_x + q_V(x) +\Lambda\,\vert\,
q_V(x)\in V\,\right\}.
\label{2.28}\ee
Due to the nondegeneracy condition  (\ref{2.22}) and the grading
assumptions, the action of $\N$ on $Q$ is a free action and
the following result holds.

\medskip
\noindent{\bf Lemma 2.}
{\em
The submanifold $Q_V\subset Q$ is a global cross section
of the gauge orbits defining  a one-to-one model of $Q/\N$.
When regarded as functions on $Q$,
the components of $q_V(x)=q_V(q(x))$ are differential polynomials, which
thus provide a free generating set of the ring 
of gauge invariant differential polynomials on $Q$.}
\medskip

Since this  lemma also  goes back to \cite{DS}  (see also \cite{rep}),
$Q_V$ is referred to as a {\em DS gauge.} 
To construct a hierarchy on $Q/\N$, we first 
 exhibit commuting vector
fields on $Q$  by means of the dressing procedure   
(recall Remark 1). 
That is  for any $b\in {\cal C}(\Lambda)$ in (\ref{2.4}),
we define  commuting vector fields
$\pa \over \pa t_b$ on $Q$, similarly to (\ref{2.14}),   by 
\be
{\pa   \over \pa t_b } q:= \left[ \R(B_b(q)), \L\right]
=\left[ \P_\alpha (B_b(q )), \L\right]
=-\left[\P_\beta(B_b(q)), \L\right]
\quad\hbox{at}\quad \L=(\pa_x + q +\Lambda)\in Q.
\label{2.29}\ee
Here $B_b(q)$ is obtained from (\ref{2.9}) 
 and the splitting (\ref{2.24})
is used to define $\R$ by (\ref{2.12}), (\ref{2.13}).
The conditions in (\ref{2.23}), (\ref{2.24}) ensure that 
(\ref{2.29}) gives a consistent evolution equation on $Q$.

The evolution equation defined by (\ref{2.29}) has a gauge 
invariant meaning.
Algebraically speaking, this means that $\pa \over \pa t_b$ induces 
a derivation of the ring 
of $\N$-invariant  differential polynomials  in $q$.
The corresponding geometric statement is that the vector field 
${\pa \over \pa t_b}$
on $Q$ can be consistently projected on $Q/\N$.
The  projectability of ${\pa \over \pa t_b}$ follows from  
the uniqueness property of the  formal dressing procedure 
stated by equation (\ref{2.8}). This  leads to the equality   
$B_b(e^\gamma \L e^{-\gamma}))=e^\gamma B_b(\L )e^{-\gamma}$, 
where $\gamma(x) \in \A_0^{<0}$ parametrizes a  gauge transformation
and we put $B_b(\L):=B_b(q)$.
Using this, it is in fact straightforward to 
show the projectability of  
${\pa \over \pa t_b}$.

If we use $Q_V$ as the model of $Q/\N$ and let  
$\pi: Q\rightarrow Q_V$ denote the natural
mapping, then the projected vector field 
 $\pi_*\left({\pa \over \pa t_b}\right)$ on $Q_V$ is described by 
an equation of the form 
\be
\pi_*\left({\pa \over \pa t_b}\right) q_V =
\left[ \R (B_b(q_V ))+\eta_b(q_V) , \pa_x + q_V +\Lambda \right],
\label{2.30}\ee
where $\eta_b(q_V(x))\in \A_0^{<0}$ is a uniquely determined  differential
polynomial in $q_V(x)$.
These vector fields 
generate the commuting flows of the KdV type hierarchy on $Q_V=Q/\N$.

To deal with the hamiltonian formalism, 
notice that $Q$ in (\ref{2.25}) 
is a Poisson submanifold of $\M$ in (\ref{2.16}) 
with respect to the Poisson bracket $\{\ ,\ \}_\R$ in (\ref{2.18}).
This  
is easy to verify by combining (\ref{2.20}) with 
(\ref{2.23}), (\ref{2.24}).
It then follows immediately that the derivative of a local
functional $f$ on $Q$ with respect to the vector field 
$\pa\over \pa t_b$ on $Q$ is given by 
\be
{\pa  \over \pa t_b} f =\{ f, \H_b\}_\R,
\label{2.31}\ee
where $\H_b$ is obtained from (\ref{2.10}). 
By the projectability of ${\pa  \over \pa t_b}$,
we know that the right hand side of (\ref{2.31})  must be 
gauge invariant  if $f$ is gauge invariant.
Since $\H_b$ is a gauge invariant local functional  on $Q$,  by 
the uniqueness property (\ref{2.8}), 
we are naturally lead to suspect that 
the Poisson bracket $\{f,g\}_\R$ of any two gauge invariant
local functionals  on $Q$ is again gauge invariant.
Indeed, under an infinitesimal gauge transformation
$\delta_\gamma \L=[\gamma, \L]$ with $\gamma(x)\in \A_0^{<0}$,
one finds for any two local functionals  $f$, $g$ on $Q$ that
\be
\delta_\gamma \{ f, g\}_\R =
\int dx\, 
\left\langle  {\delta g\over \delta q}, [\gamma_f, \L] \right\rangle 
-
\left\langle  {\delta f\over \delta q}, [\gamma_g, \L] \right\rangle 
\label{2.32}\ee
with
\be
\gamma_f=[\gamma, \R{\delta f\over\delta q}]-\
\R[\gamma,{\delta f\over\delta q}],
\qquad 
\gamma_g=[\gamma,\R{\delta g\over\delta q}]-
\R[\gamma,{\delta g\over\delta q}].
\label{2.33}\ee
Inspection shows that $\gamma_f(x), \gamma_g(x)\in \A_0^{<0}$.
Hence $\delta_\gamma \{ f, g\}_\R$ vanishes if  $f$ and $g$ are 
gauge invariant,  proving  that the Poisson bracket of gauge 
invariant local functionals  is  gauge 
invariant.

To summarize the outcome of the above, 
a  Poisson bracket is defined on $Q/\N$ by 
identifying  the local functionals on $Q/\N$ with 
the gauge invariant local functionals on $Q$ and 
determining  the Poisson bracket on these functionals
by (\ref{2.18}).
The KdV type hierarchy on $Q/\N$ is generated by the 
commuting Hamiltonians  $\H_b$ with respect to this 
induced Poisson bracket. 

In the standard case, for which $\A^0\subset \A_0$
in addition to (\ref{2.23}), one can show that  
our induced  Poisson bracket on $Q$ 
coincides with the `second' Poisson bracket given in \cite{Prin2}.
In this case one also has an alternative
interpretation of the `second' Poisson bracket on $Q/\N$ 
based on the r-matrix associated with the splitting $\A=\A_{\geq 0}
+\A_{ <0}$ (see e.g.~\cite{Marseille}).
However, in the case of a splitting in 
(\ref{2.24}) for which $\beta^0\neq \{ 0\}$ this second
r-matrix does not even lead to consistent flows on $Q$, 
and hence we do not have such an alternative interpretation.

\medskip

Given the data $(\A, \Lambda, d_1, d_0; \alpha^0, \beta^0)$ with which
we associated a KdV  type system,
we also have the modified KdV type system
corresponding to the data $(\A, \Lambda, d_1; \alpha^0, \beta^0)$.
This modified KdV system can be restricted to the subspace of its
phase space $\Theta$ in (\ref{2.3}) given by 
\be
\Xi := \Theta\cap Q  =\left\{ \L=\pa_x + \xi(x) +\Lambda\,\vert\,
\xi(x)\in  \A^{<k} \cap \A^{\geq 0}\cap \A_{\geq 0}  \,\right\}.
\label{Xi}\ee
In fact,   $\Xi\subset \Theta\subset Q\subset \M$ is 
in this case a chain of  
Poisson submanifolds with respect to $\{\ ,\ \}_\R$ in (\ref{2.18}). 
One can  further check that 
\be
{\mathrm dim}\left(  \A^{<k} \cap \A^{\geq 0}\cap \A_{\geq 0}  \right)
= {\mathrm dim}\left( \A^{<k } \cap \A_{\geq 0}\right) -
{\mathrm dim}\left( \A_{0}^{ <0}\right),
\label{2.35}\ee 
which means that the number of the components of $\xi(x)$ coincides
with the number of the independent  KdV fields.
The map 
\be
\mu : \Xi \rightarrow Q/\N
\label{Miura}\ee 
induced by the natural
projection $Q\rightarrow Q/\N$ 
is a generalization of the well-known Miura map.
This maps converts the modified KdV type flows on $\Xi$ to the KdV type flows 
on $Q/\N$.
In the hamiltonian setting, $\mu$ is a Poisson map with respect
to the (linear) Poisson bracket $\{\ ,\ \}_\R$ on $\Xi$ and the 
(nonlinear) Poisson bracket on $Q/\N$ obtained as  reduction
of the Poisson bracket $\{\ ,\ \}_\R$ on $Q$.

\medskip
\noindent
{\em Remark 2.}
One can naturally extend  the action of $\N$ in (\ref{2.26})
 to the manifold $\M$ in (\ref{2.16}).
Although (like the action on $Q$) 
this action of $\N$ on $\M$ does not leave the Poisson bracket 
$\{\ ,\ \}_\R$ in (\ref{2.18}) invariant,
it is an `admissible' action in the sense that the
$\N$-invariant functions close under the Poisson bracket.
This property permits  to consider hamiltonian symmetry  reduction 
of the hierarchy on $\M$, mentioned in Remark 1, with respect to $\N$.
Then  $Q/\N \subset \M/\N$ can be interpreted as a Poisson submanifold
and the KdV type hierarchy as a subsystem of the reduced 
hierarchy on $\M/\N$.

\section{Application to  nonstandard Gelfand-Dickey hierarchies}
\setcounter{equation}{0}

In fact, our original motivation for this work was to understand whether
the nonstandard Gelfand-Dickey hierarchy \cite{kuper,OS,Oevel2} 
defined by (\ref{1.2})
can be interpreted in the Drinfeld-Sokolov framework.
We here explain that it can indeed 
be obtained as an example of the general construction 
presented in the previous section.
It turns out that one needs to use a nontrivial splitting
with $\beta^0\neq \{0\}$ in (\ref{2.24})
to derive this system.
For completeness, we also give the interpretation of the 
other nonstandard Gelfand-Dickey hierarchy \cite{OS,Oevel2} defined by 
\begin{equation}
{\partial \over \partial t_m} L= [ ( L^{m\over n+1})_{> 0}, L] 
\qquad\hbox{\rm for}\qquad
L=\left(\partial^{n} +\sum_{i=1}^{n} u_i \partial^{n-i}\right)\pa. 
\label{3.1}\end{equation}
In this  case the input data  
satisfy the standard conditions, for which 
$\A^0\subset \A_0$.

In the next subsection, we collect known results  that will 
be used in the derivation of the nonstandard Gelfand-Dickey systems
presented in subsections 3.2 and 3.3.

\subsection{Generalized KP systems}

We now recall some basic facts about the standard and 
nonstandard KP systems from which the Gelfand-Dickey systems arise by
reduction to Poisson submanifolds.
For a detailed exposition, see  e.g.~\cite{Di,Oevel2} and 
references therein.
Note that in this paper only the quadratic Poisson
brackets will be considered, even 
though these systems are known to possess linear Poisson brackets as well.

We denote by $\D$ the 
associative algebra of scalar pseudo-differential operators 
(PDOs) of the form  
\begin{equation}
L=\sum_{i=0}^{\infty}u_i\partial^{n-i}
\qquad\qquad
\forall n\in \Z.
\label{3.2}\end{equation}
The transposition $L \rightarrow L^{t}=\sum_{i=0}^{\infty}
( - \partial )^{n-i}u_i$ is an anti-involution of  $\D$.
The projectors $P_{\geq 0}$, $P_{<0}$, $P_{0}$, $P_{>0}$ and $P_{\leq 0}$ 
on the corresponding associative subalgebras of $\D$
are given by
\begin{eqnarray}
P_{\geq 0}(L)= \sum_{i=0}^{n }u_i\partial^{n-i} , \qquad 
P_{<0}(L)= \sum_{i=1}^{\infty}u_{n+i}\partial^{-i} \nonumber\\
P_{0}(L)= u_n , \qquad
 P_{>0} = P_{\geq 0}-P_{0} , 
\qquad
 P_{\leq 0} =P_{<0}+P_{0}.
\label{3.4}\end{eqnarray}
The Adler trace \cite{Ad} on $\D$ is given by 
${\mathrm{Tr}}(L)=\int dx \, {\mathrm{res}}(L)$ with 
${\mathrm{res}}(L)=u_{n+1}$. 
We shall use the linear functionals on $\D$ defined by 
$l_{X}(L)={\mathrm{Tr} }(LX)$, where $X$ is some constant PDO.
For a fixed positive integer $n$, 
we shall often consider the affine subspace  $\D_n$ of $\D$,
\be
\D_n=\{\, L= \pa^n + \sum_{i=1}^\infty u_i \pa^{n-i}\,\}.
\label{3.3}\ee
 
\medskip
\noindent {\em 3.1.1.\ The standard KP hierarchy.}
The KP hierarchy is defined 
with the aid of the splitting of $\D$
into differential and purely pseudo-differential operators,
which 
yields the antisymmetric r-matrix $R={1\over 2}(P_{\geq 0}-P_{<0})$. 
In association with $R$, there exists a one parameter family of 
{\it local} quadratic Poisson brackets
on $\D$ 
\be
\{ l_{X}, l_{Y} \}_{\mathrm{GD}}^{\nu}(L) = {\mathrm{Tr}} \left( 
LXR(LY) -XLR(YL) \right)
+ \nu \int dx (D^{-1} {\mathrm{res}}[L,X]){\mathrm{res}}[L,Y],
\label{3.5}\ee
where $\partial_x(D^{-1}f) = f$. 
Notice that the constant ambiguity in the definition of $(D^{-1} f)$
drops out from formula (\ref{3.5}), which defines 
a local Poisson bracket since 
${\mathrm{res}}[L,X]$ is a total derivative.
The `second' Adler-Gelfand-Dickey bracket \cite{Di} corresponds to $\nu =0$.
The possibility to add the term proportional with $\nu$ on the
right hand side of (\ref{3.5}) was apparently first noticed in \cite{DIZ}.
 For any  $n>0$, $\D_n\subset \D$ 
is a Poisson submanifold 
with respect to the family of brackets in (\ref{3.5}).
For any complex number $c$, 
the local (differential polynomial) map $F_c$ given on $\D_n$ by 
\be
F_c(L)=e^{c\Phi}Le^{-c\Phi},
\qquad
 \Phi=D^{-1}u_1, 
\ee
which is invertible except for $c = {1\over n}$,
is a Poisson map according to  
\be
\{l_X\circ F_c,l_Y\circ F_c\}^\nu_{\mathrm{GD}}
=\{l_X,l_Y\}^{\nu_c}_{\mathrm{GD}}\circ F_c,\qquad
\nu_c=\nu+c(2-nc)(1-n\nu).
\ee
Note that $\nu=\frac{1}{n}$ is a fixed point for this transformation.
The operators $L\in \D_n$ satisfying the condition 
$L=P_{\geq 0}(L)$ form a Poisson submanifold with respect to
$\{\ ,\ \}^\nu_{\mathrm{GD}}$ for any $\nu$,
whereas those of the form 
\begin{equation}
L= \partial^{n}+\sum_{i=2}^{n}u_i\partial^{n-i}
\label{3.7}\end{equation}
form a Poisson submanifold for the value $\nu= {1\over n}$ only.

On $\D_n$, 
the commuting KP flows\begin{footnote}
{Originally, only the hierarchy with the value $n=1$ was called the KP 
hierarchy.}
\end{footnote}
 \begin{equation}
{\partial\over\partial t_m}L=[R(L^{m\over n}),L]
=[(L^{m\over n})_{\geq 0}, L]
\qquad m=1,2, \dots 
\label{3.6}\end{equation}
are generated by the Hamiltonians 
$H_{m}(L)={n \over m}{\mathrm{Tr}}(L^{{m \over n}})$ with respect to
any of the brackets in (\ref{3.5}).
Restriction of the KP flows (\ref{3.6}) to (\ref{3.7})
gives the standard $n^{th}$ KdV hierarchy.

Let us now write the operator 
$L=P_{\geq 0}(L)\in \D_n$ 
in the factorized  form
\begin{equation}
L=\partial^{n}+\sum_{i=1}^{n}u_i\partial^{n-i} = 
(\partial+\xi_n)(\partial+\xi_{n-1})\cdots (\partial+\xi_1)
\label{3.8}\end{equation}
which yields the Miura transformation, each field $u_i$ being expressed 
as a differential polynomial in the $\xi_i$. 
Then the generalization in \cite{Liu} of the 
Kupershmidt-Wilson theorem \cite{kuwi} asserts that 
the  quadratic bracket $\{ f, g \}_{\mathrm{GD}}^{\nu}(L)$ is 
equal to the bracket
\begin{equation}
\int dx\, \sum_{i,l=1}^{n} \left( \frac{\delta f}{\delta \xi_i}\right)_{x} 
(\delta_{il} -\nu) \left( \frac{\delta g}{\delta \xi_l} \right)
\label{3.9}\end{equation}
when the ${u_i}$ and the ${\xi_i}$ are related through the Miura 
transformation.
Notice that this bracket is invariant under any permutation of the 
$\xi_{i}$ and under a global change of sign of them. 

\medskip
\noindent {\em 3.1.2.\ The nonstandard KP hierarchy.}
The relevance of `nonstandard' splittings of $\D$ to soliton equations 
was apparently first noticed in \cite{reyman}, 
see also \cite{kuper,kiso}. 
The definition of the nonstandard KP hierarchy of our interest 
is based on the splitting of 
$\D$ into purely differential operators  and pseudo-differential 
operators containing the constant term. 
This splitting gives rise to the non-antisymmetric r-matrix 
\be
{\hat R}={1\over 2}(P_{>0}-P_{\leq 0}).
\label{3.10}\ee 
In correspondence with this r-matrix,
two quadratic {\em local} Poisson brackets  
have been defined on $\D$  in  \cite{OS,Oevel2}:
\begin{eqnarray}
&\{ l_{X}, l_{Y} \}_{\mathrm{O}}^{A}(L) =
{\mathrm{Tr}}(LX{\hat R}(LY) - XL{\hat R}(YL))&
\qquad\qquad \qquad
\nonumber \\
&\phantom{\{ l_{X}, l_{Y} \}_{\mathrm{O}}^{A}(L) AAAAA }
 +\mbox{\rm Tr}([L,Y]_0 XL + [L,Y] (LX)_0 + 
(D^{-1} {\mathrm{res}}[L,Y]) [X,L]),&
\label{braca}\end{eqnarray} 
\begin{eqnarray}
&\{ l_{X}, l_{Y} \}_{\mathrm{O}}^{B}(L) =
{\mathrm{Tr}}(LX{\hat R}(LY) - XL{\hat R}(YL))&
\qquad\qquad \qquad
\nonumber \\
&\phantom{\{ l_{X}, l_{Y} \}_{\mathrm{O}}^{A}(L) AAAAA }
 +{\mathrm{Tr}}([L,Y]_0 LX + [L,Y] (XL)_0 - (D^{-1} 
{\mathrm{res}}[L,Y]) [X,L]).&
\label{bracb}\end{eqnarray}
For any fixed $n>0$, $\D_n\subset \D$ is a Poisson subspace
with respect to both of the above Poisson brackets. 
The two brackets admit different 
Poisson subspaces with a finite number of fields. 
In the case of bracket (\ref{braca}), the operators $L^A$
of the form 
\be
L^{A}=\pa^{n-1} +\sum_{i=1}^{n-1} u_i \pa^{n-1-i}+ \pa^{-1} u_{n}
\label{LA}\ee
form a Poisson submanifold for any $n\geq 1$.
In the case of bracket 
(\ref{bracb}), 
the operators 
\be
L^{B}=(\pa^n + \sum_{i=1}^n u_i \pa^{n-i})\pa
\label{LB}\ee
form a Poisson submanifold for any $n\geq 1$.

On the space $\D_n$,
the nonstandard KP 
hierarchy\footnote{The term `modified' KP
hierarchy  is also used in the literature,
especially in the $n=1$ case \cite{kuper2,satu}.}
is given by the commuting flows  
\begin{equation}
{\partial\over\partial t_m}L=[{\hat R}(L^{m\over n}),L]=
[(L^{m\over n})_{>0}, L]
\qquad m=1,2,\ldots\,.
\label{nsKP}\end{equation}
These flows are  generated by the 
Hamiltonians $H_{m}={n \over m}{\mathrm{Tr}}(L^{{m \over n}})$
with respect to any of the brackets (\ref{braca}), (\ref{bracb}).
In contrast to the standard case, the coefficient
$u_1$ of the subleading term of $L$ is now dynamical, 
but $H_0:=\int dx\, u_1$ is still constant with respect to the 
flows.
Restriction of these flows to the sets of operators $\{ L^A\}$ in (\ref{LA})
and $\{ L^B\}$ in (\ref{LB}) 
yields the nonstandard Gelfand-Dickey hierarchies 
in (\ref{1.2}) and in (\ref{3.1}), respectively.
In the case (A) one obtains nontrivial flows for $n\geq 2$ only.

There exists a Poisson equivalence between the   
brackets $\{\ ,\ \}_{\mathrm{O}}^{A,B}$ 
and  $\{\ ,\ \}_{\mathrm{GD}}^\nu$
with $\nu=\pm 1$:
\begin{eqnarray}
\{ l_{X}\circ p_A  , l_{Y} \circ p_A\}_{\mathrm{GD}}^{+1} 
= \{ l_{X} , l_{Y} \}_{\mathrm{O}}^{A}\circ p_A 
\label{Poissona}
\\
\{ l_{X}\circ p_B , l_{Y}\circ p_B \}_{\mathrm{GD}}^{-1} = 
\{ l_{X} , l_{Y} \}_{\mathrm{O}}^{B} \circ p_B 
\label{Poissonb}
\end{eqnarray}
where $p_A$, $p_B$ are two invertible maps 
on $\D$ defined by $p_A(L)=\partial^{-1}L$ and 
$p_B(L)=L \partial$. 
This result, given in \cite{Huang,Liu} (see also \cite{GD}),
allows to derive the properties
of the brackets $\{\ ,\ \}_{\mathrm{O}}^{A,B}$  from familiar 
properties of the bracket $\{\ ,\ \}^\nu_{\mathrm{GD}}$.
In particular, it 
allows for a straightforward derivation \cite{Liu,Shaw} 
of generalized Kupershmidt-Wilson theorems for these brackets.
For this purpose,  one writes the operators 
$L^{A}\in p_A(P_{\geq 0}(\D_n))$ in (\ref{LA}) 
and $L^{B}\in p_B(P_{\geq 0}(\D_n))$ in (\ref{LB})
in a multiplicative form as 
\begin{eqnarray}
&L^{A}=
\partial^{-1}(\partial+\xi_n)(\partial+\xi_{n-1})\cdots(\partial+\xi_1)
\label{Miuraa}
\\
&L^{B}=
(\partial+\xi_n)(\partial+\xi_{n-1})\cdots (\partial+\xi_1)\partial. 
\label{Miurab}
\end{eqnarray} 
Then the Kupershmidt-Wilson  theorem for $\{\ ,\ \}_{\mathrm GD}^\nu$
with $\nu = \pm 1$ and the relations 
(\ref{Poissona}), (\ref{Poissonb}) imply that 
if the $u_i$ are expressed through the $\xi_i$ by the   
 Miura transformations (\ref{Miuraa}), (\ref{Miurab}), then 
the brackets $\{\ ,\ \}_{\mathrm{O}}^{A,B}$   on $\{ L^{A,B}\}$ 
are equal to 
\begin{equation}
\int dx\, \sum_{i,l=1}^{n} \left( \frac{\delta f}{\delta \xi_i}
\right)_{x}(\delta_{il}-\nu^{A,B}) 
\left( \frac{\delta g}{\delta \xi_l}  \right)
\label{PBab}
\end{equation}
with $\nu^{A}=-\nu^{B}=1$, respectively.
These Miura  transformations will be quite useful for us.

Although they intertwine the Poisson brackets, 
neither $p_A$ or $p_B$ converts   
the standard Gelfand-Dickey  hierarchy into  the 
nonstandard one since the 
commuting Hamiltonians are not intertwined by these maps.

\medskip
\noindent
{\em Remark 3.}
It is well-known \cite{STS} that the quadratic
Adler-Gelfand-Dickey bracket is a version of the Sklyanin
bracket and its Jacobi identity depends on the 
antisymmetry of the r-matrix $R={1\over 2}(P_{\geq 0} - P_{<0})$.
General results on quadratic Poisson brackets on Lie groups
associated with non-antisymmetric r-matrices 
have been obtained in \cite{Maillet}, 
and equivalent results are found in \cite{S} in the context 
of associative algebras,
see also \cite{LP, OR} which deal with  special cases.
The brackets $\{\ ,\ \}_{\mathrm GD}^\nu$ as well as the brackets 
$\{\ ,\ \}_{\mathrm{O}}^{A,B}$ are identified in \cite{GD} as special
cases of the `$(a,b,c,d)$-scheme' of \cite{Maillet}, 
with nonlocal operators $a$, $b$, $c$, $d$. 
For the brackets $\{\ ,\ \}_{\mathrm{O}}^{A,B}$ an
equivalent identification in terms of the
notation of \cite{S} is contained in \cite{Oevel2}. 

\subsection{Nonstandard Gelfand-Dickey system of type A}

Now we show that the nonstandard Gelfand-Dickey system 
in (\ref{1.2}) can be recovered within the generalized Drinfeld-Sokolov 
formalism developed in section 2  with an appropriate choice of the 
sextuplet $({\cal A},\Lambda, d_1,d_0;\alpha^0,\beta^0)$.
Our demonstration below will be purely deductive; 
the right choice of data was originally found by a long explicit
inspection of the linear problem for $L$ in (\ref{1.2}), which involved 
some guesswork too.

We consider the algebra ${\cal A}=sl(n)\otimes\C[\lambda,\lambda^{-1}]$. 
We denote by  $e_{i,j}$ the 
$n\times n$ matrix with $1$ at 
the intersection of line $i$ and column $j$ and $0$ everywhere else, and
introduce the two compatible gradations of ${\cal A}$
\begin{eqnarray}
&d_1 = (n-1)\lambda \partial_{\lambda} +
{\mathrm{ad\,}}\K_A, \quad 
\K_A = \sum_{k=1}^{n}{n+1-2k\over 2}\, e_{k,k}
\label{3.23}\\
&d_0 = \lambda \partial_{\lambda}.
\label{nsgradinga}\end{eqnarray}
Note that $d_0$ is the homogeneous gradation like in the derivation  
of the standard system (\ref{1.1}) \cite{DS}, but $d_1$ is not the  principal 
gradation. These gradations satisfy the conditions in (\ref{2.23}), 
but  do not satisfy the condition  $\A^0 \subset \A_0$. 
Then for $n \geq 3$ we choose\begin{footnote}{The  $n=2$ case is special
and will be treated separately.}\end{footnote}
\begin{equation}
\Lambda := \sum_{k=1}^{n-1}e_{k,k+1}+\lambda (e_{n-1,1}+e_{n,2}),
\label{nslambda}\end{equation}
which has $d_1$-grade one.
One can check that $\Lambda$ is a regular semisimple element, 
that is to say  
${\cal A} = \ker \oplus \im$, and 
$\ker \subset \A$ is an abelian subalgebra.
The vector space $\ker$ is generated by the 
homogeneous elements $\Lambda_m\in \A^m$ given by 
\begin{equation}
  \Lambda_{l(n-1)+r} =
  (2\lambda)^{l}(\Lambda^{r}-\delta_{r,n-1}{2(n-1)\lambda \over n}I_{n})
\qquad\hbox{with}\qquad
 1\leq r \leq n-1, \quad l \in \Z.
\label{3.26}\end{equation}
The splitting of $\A^0$  of the form in (\ref{2.24}) is in this case 
defined by 
\be
\beta^{0} :=( \ker)^{0}={\mathrm{span}}\{ \Lambda_0\}.
\label{3.27}\ee

\medskip
\noindent {\em 3.2.1.~The modified KdV type system.}
We shall first identify the  mod-KdV type system associated 
with the above  sextuplet as the modified 
nonstandard Gelfand-Dickey system of  type A based on the 
factorized Lax operator in (\ref{Miuraa}). 
An element  of  the mod-KdV phase space  $\Xi$ in (\ref{Xi})
can now  be parametrized by 
\begin{equation}
\xi (x) = \sum_{k=1}^{n}(\xi_{k}-\sigma)e_{k,k} +
 \lambda(\xi_{1}+\xi_{n})e_{n,1},\qquad
\sigma = { 1 \over n}\sum_{k=1}^{n}\xi_{k}.
\end{equation}
The explicit evaluation of the Poisson bracket (\ref{2.18})
on $\Xi$ yields exactly the bracket in (\ref{PBab})
that corresponds to the Miura map for 
the bracket $\{\ ,\ \}_{\mathrm{O}}^A$. 
The identification of the conserved quantities  can be performed
 through the linear problem and the elimination procedure 
following, e.g., the lines of \cite{FM}. 
If $\Psi = (\psi_1,\psi_2,\ldots,\psi_n)^{t}$ then one gets from the 
linear problem ${\cal L}\Psi =0$ the eigenvalue equation 
\begin{equation}
(\partial -\sigma)^{-1}(\partial+\xi_n -\sigma)
(\partial+\xi_{n-1} -\sigma)
\cdots(\partial+\xi_1 -\sigma) \psi_{1}= 2 (-1)^{n-1} \lambda \psi_{1},
\end{equation}
which is equivalent to 
\begin{equation}
L \tilde{\psi}_{1} = 
\partial^{-1}(\partial+\xi_n)
(\partial+\xi_{n-1})\cdots(\partial+\xi_1) \tilde{\psi}_{1}= 
2 (-1)^{n-1} \lambda \tilde{\psi}_{1}
\quad \hbox{with}\quad   \tilde{\psi}_{1} :=e^{-(D^{-1}\sigma)}\psi_1.
\label{eigen1}
\end{equation}
According to subsection 2.1,
the commuting Hamiltonians of the mod-KdV type system can be chosen as  
$\H_{k}=\int dx\, h_{k}(x)$ where  
$e^{{\mathrm{ad}} F}(\L) = \pa_x + \Lambda +
\sum_{m=0}^{\infty}h_{m}(x)\Lambda_{-m}$
is defined by (\ref{2.6}).
Then $\tilde{\psi}_{1}$ may then be computed in two different ways, from 
${\cal L}\Psi =0$ and from (\ref{eigen1}), along the lines of \cite{FM}. 
By comparison of the two results, one obtains 
\begin{eqnarray}
& \H_{0}= {1 \over n-1}\int dx( \sum_{k=1}^{n}\xi_k) \nonumber &\\
& \H_{m}=-{a_m \over m} (-1)^m {\mathrm{Tr}}(L^{m \over n-1})   
\quad\hbox{for}\quad m \geq 1, &
\label{3.28}\end{eqnarray}
with $a_m =n$ if $m$ is zero modulo $(n-1)$, and $a_m =1$ in 
the other cases. 
Except for normalization, the Hamiltonians $\H_m$
for $m\geq 1$ are thus identical with the $H_m$ that generate 
the flows in (\ref{1.2}) and $\H_0$ gives the conserved charge
$H_0$ mentioned 
after (\ref{nsKP}).
This  proves the equivalence of the  two modified systems.

\medskip
\noindent {\em 3.2.2.~The KdV type system.}
We shall now identify the generalized KdV type 
system that results from the construction of subsection 2.2 
using the above sextuplet as the nonstandard Gelfand-Dickey system 
in (\ref{1.2}).
The nondegeneracy condition (\ref{2.22}) holds in our case 
since $\Lambda$ is a regular element.
We can parametrize  the phase space $Q/\N$ of the KdV system by the DS 
gauge slice $Q_{V}$ in (\ref{2.28}) whose general element is written as 
\begin{equation}
q_{V}(x)=2\lambda v_{1}e_{n,1}+ \sum_{k=1}^{n-1}(-1)^{n-k}v_{n+1-k}e_{n,k}.
\end{equation}
{}From the linear problem ${\cal L}\Psi =0$ for $\L\in Q$ (\ref{2.25}),  
we obtain a gauge invariant eigenvalue equation on $\psi_{1}$,
which in the DS gauge becomes 
\begin{equation}
(\partial -v_{1})^{-1}(\partial^{n}+ 
\sum_{k=2}^{n}v_{k}\partial^{n-k}){\psi}_{1} = 
2(-1)^{n-1} \lambda {\psi}_{1},
\end{equation}
or equivalently
\begin{equation}
L\tilde{\psi}_1=
(\partial^{n-1}+\sum_{k=1}^{n-1}u_{k}\partial^{n-1-k}+
\partial^{-1}u_{n})\tilde{\psi}_{1} = 2(-1)^{n-1} \lambda \tilde{\psi}_{1},
\qquad
\tilde{\psi}_{1}= e^{-(D^{-1}v_1)}\psi_{1}.
\label{eigen2}
\end{equation}
The $u_{k}$ have differential polynomial expressions in terms of 
the $v_{k}$. These relations are invertible and the $v_{k}$ are 
also differential polynomials of the $u_{k}$. 
Hence the points of $Q_{V}$ may be  parametrized by the functions $u_{k}$.
The correspondence between the two parametrization of $L$ in 
(\ref{eigen1}) and in (\ref{eigen2}) provides a model for the 
Miura map $\mu: \Xi \rightarrow Q/\N=Q_V$ in (\ref{Miura}), which 
is a Poisson map when $\Xi$ is equipped with the
 linear bracket (2.18) and $Q/{\cal N} = Q_{V}$ 
is equipped with the nonlinear bracket obtained as the reduction of 
the linear bracket on $Q$. 
Since this map is given by just the factorization of $L$, using 
the identification between the linear bracket on $\Xi$ with the
bracket in (\ref{PBab}) that appears in   
 the generalized Kupershmidt-Wilson theorem 
for the bracket $\{\ ,\ \}_{\mathrm{O}}^A$ (\ref{braca}),
 we conclude that the nonlinear bracket on $Q/{\cal N} =Q_{V}$ 
coincides with the bracket $\{\ ,\ \}_{\mathrm{O}}^A$  on the set 
of the Lax operators $L=L^A$ in (\ref{LA}).
The identification  between the respective sets of commuting 
Hamiltonians has already been established in (\ref{3.28}).

\medskip
\noindent {\em 3.2.3.~The case $n=2$.}
This case is slightly different from the generic one because of 
the form of $\Lambda$, now we choose $\Lambda= e_{1,2}+\lambda^2e_{2,1}$. 
The element $\Lambda$ is semisimple and regular with 
$\ker = 
{\mathrm{span}}\{ \lambda^k \Lambda\,\vert\, k \in \Z\}$.
Apart from this, the sextuplet $({\cal A},\Lambda, 
d_1,d_0;\alpha^0,\beta^0)$ does not differ from the generic case.
We briefly describe the identification of the associated 
KdV type system with the nonstandard Gelfand-Dickey system in (\ref{1.2}) 
for $n=2$.
A convenient Drinfeld-Sokolov gauge is defined by
\begin{equation}
q_{V}(x)=(-\lambda u_{1}-{1\over 4}(4u_{2}+2u_{1x} -u_{1}^{2}))e_{2,1}.
\end{equation}
{}From the linear problem ${\cal L}\Psi =0$,  
we obtain the  eigenvalue equation on $\psi_{1}$
\begin{equation}
(\partial^2+u_{2}+{u_{1x} \over 2}-{u_{1}^{2} \over 4}+
\lambda u_{1})\psi_{1} = \lambda^2 \psi_{1},
\end{equation}
or equivalently 
\begin{equation}
L\tilde\psi_1=(\partial+u_{1}+\partial^{-1}u_{2})\tilde{\psi}_{1} 
= 2\lambda\tilde{\psi}_{1},\qquad
\tilde{\psi}_{1}= \exp({-\lambda x-{1\over 2}
D^{-1}{ u_1}})\psi_{1}.
\label{eigen5}
\end{equation}
Thanks to the low number of fields, the Poisson brackets of the $u$'s 
may easily  be computed by reduction of (\ref{2.18}), and coincide 
with those obtained from the bracket in  (\ref{braca}). 
The identification of the commuting Hamiltonians  
is obtained as in the other cases.

\medskip
\noindent
{\em Remark 4.}
Another example for which the technique developed in 
section 2 applies is the following.
The algebra ${\cal A}$ in the sextuplet 
$({\cal A},\Lambda, d_1, d_0 ;\alpha^0,\beta^0)$ is now 
${\cal A}=sl(n+1)\otimes\C[\lambda,\lambda^{-1}]$, but we keep the same 
matrix $\Lambda$ of equation \p{nslambda}, and the same gradations
(\ref{3.23},\ref{nsgradinga}) as before. 
Clearly, $\Lambda$ still is a semisimple element. 
A basis for $( \ker)^{0}$ is now given by the two elements 
$\Lambda_0$ in \p{3.26} and 
$\Omega=e_{n+1,n+1}-{1\over n}\sum^n_{k=1}e_{k,k}$, and we  take
\be
\beta^{0} :={\mathrm{span}}\{ \Lambda_0\}.
\ee
A Drinfeld-Sokolov gauge may be parametrized by 
\begin{equation}
q_{V}(x)=2\lambda (v_{1}-{\vartheta\over n})e_{n,1}+ 
\sum_{k=1}^{n-1}(-1)^{n-k}v_{n+1-k}e_{n,k}
+\vartheta \Omega +(-1)^n\varphi e_{n,n+1}+\chi e_{n+1,1}.
\end{equation}
Then the standard elimination procedure leads from the linear problem
 ${\cal L}\Psi =0$ to the  eigenvalue equation on $\psi_{1}$
\begin{equation}
(\partial -v_{1})^{-1}((\partial-{\vartheta\over n})^{n}+ 
\sum_{k=2}^{n}v_{k}(\partial-{\vartheta\over n})^{n-k}
+\varphi(\partial -\vartheta)^{-1}\chi){\psi}_{1} = 
2(-1)^{n-1} \lambda {\psi}_{1},
\end{equation}
or equivalently, with $\tilde{\psi}_{1}= e^{-(D^{-1}v_1)}\psi_{1}$,
\begin{equation}
L\tilde{\psi}_1=
(\partial^{n-1}+\sum_{k=1}^{n-1}u_{k}\partial^{n-1-k}+
\partial^{-1}u_{n}+\partial^{-1}\varphi(\partial +w)^{-1}
\chi)\tilde{\psi}_{1} = 2(-1)^{n-1} \lambda \tilde{\psi}_{1}.
\label{eigen20}
\end{equation}
The fields $u_{k}$ and $w$ have invertible differential polynomial 
expressions in terms of the $v_{k}$ and $\vartheta$.  
It is not hard to check that the nonstandard KP equations \p{nsKP} define 
consistent flows for the operator $L$ in  \p{eigen20}. 
Moreover, the field $w$ 
does not evolve under these flows, and may be set to zero. Then, if one 
introduces a field $\Phi$ which is a primitive of $\varphi$, 
$\Phi=(D^{-1} \varphi)$, the Lax operator $L $ may be brought to the form
\be 
L=\partial^{n-1}+\sum_{k=1}^{n-1}u_{k}\partial^{n-1-k}+
\partial^{-1}(u_{n}-\Phi\chi)+\Phi\partial^{-1}\chi, \ee
which is one of the Poisson subspaces of the bracket  \p{braca} given 
in \cite{Oevel2}. 
The restricted Poisson bracket is nonlocal in this parametrization
\cite{Oevel2}.
One should note that the set of Lax operators $L$ in \p{eigen20}
also defines a Poisson subspace of the bracket \p{braca}, 
which may be extended to a {\em local} bracket on the set of fields
$\{u_k,w,\varphi,\chi\}$.
Finally, 
it is clear that using the natural 
embedding of $sl(n)$ into $sl(n+m)$ for any $m\geq 1$
while keeping the same element $\Lambda$, one would reach a Lax operator
containing $m$-component vector versions of $\varphi$ and $\chi$.
More generally, constrained nonstandard matrix KP systems 
having  Lax operators similar in form to $L$
in \p{eigen20} can also be derived by a slight modification
of this example.

\medskip
\noindent
{\em Remark 5.}
Restricting to the case $n=2l+1$, we notice that the element $\Lambda$ in
\p{nslambda} may be conjugated to 
\be
\hat\Lambda=\sum_{i=1}^{l}e_{i,i+1}-\sum_{i=l+1}^{2l}e_{i,i+1}
-\lambda e_{2l,1} +\lambda e_{2l+1,2},  \ee
which is thus semisimple and has $d_1$-grade one. 
We then consider the involution $\zeta$ of the algebra ${\cal A}$ defined 
on some element $X$ by
\be 
X\mapsto  \zeta(X)=-\eta X^t\eta,\qquad 
\eta=\sum_{i=1}^{2l+1}e_{i,2l+2-i}. \ee
The elements in ${\cal A}$ invariant under this involution form the 
loop algebra of $so(2l+1)$ and  $\hat \Lambda$ is an invariant element. 
The gradation $d_1$ in \p{3.23} and the 
homogeneous gradation $d_0$ both commute with the involution $\zeta$. 
Moreover, the r-matrix ${\cal R}$ in \p{2.13} with 
$\alpha_0$ as in \p{2.24} and $\beta_0$ as in \p{3.27} also commutes with 
$\zeta$. 
Therefore, using an invariant element $b\in{\cal C}({\Lambda})$, 
one may restrict the flow  \p{2.29} to those elements $q$ which are 
invariant under $\zeta$. 
The map $\P_{\beta_0}$ is identically zero on the invariant subalgebra. 
As a consequence, the restricted nonstandard hierarchy is 
just the standard Drinfeld-Sokolov hierarchy \cite{DS} based on  
the algebra $so(2l+1)\otimes\C[\lambda,\lambda^{-1}]$,
whose Lax operator satisfies $\pa^{-1} L^t \pa = L$.
If $n=2l$, then a reduction of (\ref{1.2}) to Lax operators satisfying 
$\pa^{-1} L^t \pa =-L$ is possible \cite{kuper}, but in this case 
we do not have an interpretation in the Drinfeld-Sokolov approach
at present.

\subsection{Nonstandard Gelfand-Dickey system of type B}

The nonstandard Gelfand-Dickey hierarchy 
of the type in (\ref{3.1}) can be recovered 
within the usual Drinfeld-Sokolov formalism \cite{Prin1,Prin2} with 
an appropriate choice of the quadruplet $({\cal A}, \Lambda, d_1, d_0)$.

We consider ${\cal A}=sl(n)\otimes\C[\lambda,\lambda^{-1}]$ 
endowed with the two compatible gradations
\begin{eqnarray}
&d_1 = n\lambda \partial_{\lambda} +{\mathrm{ad\,}} \K_B, \qquad
 \K_B =\sum_{k=1}^n {n+1-2k\over 2} e_{k,k}
\\
&d_0 = 2\lambda \partial_{\lambda}+ {\mathrm{ad\,}}\K, \qquad 
\K ={1-n\over n} e_{n,n} + {1\over n} \sum_{k=1}^{n-1} e_{k,k}.
\label{gradnsb}\end{eqnarray}
Here $d_1$ is the principal gradation, but
 $d_0$ is not the homogeneous gradation.
The assumptions in (\ref{2.23}) as well as
the condition $\A^0\subset\A_0$ are satisfied.
We choose for $\Lambda$ the standard regular semisimple 
element of $d_1$-grade one: 
\begin{equation}
\Lambda = \sum_{k=1}^{n-1}e_{k,k+1}+\lambda e_{n,1}.
\end{equation}
The abelian algebra $\ker$ is generated by 
the matrices $\Lambda^{m}$ for $m$ not a multiple of $n$.

\medskip 
\noindent {\em 3.3.1.~The modified KdV type system.}
We wish to identify the mod-KdV type system defined by the above
quadruplet with the system belonging to the factorized Lax operator, 
of order $n$, of the form $L=L^B$ in (\ref{LB}). 
For this we now parametrize  
the phase space $ \Xi$ in (\ref{Xi}) by 
\begin{equation}
\xi (x) = \sum_{k=1}^{n-1}(\xi_{k}-\sigma)e_{k,k} -\sigma e_{n,n},
\qquad 
\sigma = { 1 \over n}\sum_{k=1}^{n-1}\xi_{k}.
\end{equation}
The explicit evaluation of the Poisson bracket (\ref{2.18}) 
on $\Xi$ yields exactly the bracket  given in (\ref{PBab}). 
The linear problem ${\cal L}\Psi =0$, where 
$\Psi = (\psi_1, \psi_2, \ldots, \psi_n)^{t}$,   
leads to the eigenvalue equation 
\begin{equation}
L\tilde{\psi}_{n}=(\partial+\xi_{n-1})(\partial+\xi_{n-2})\cdots 
(\partial+\xi_1) \partial \tilde{\psi}_{n}= (-1)^{n}
\lambda \tilde{\psi}_{n}
\label{eigen3}\end{equation}
with $\tilde{\psi}_{n} =e^{-(D^{-1}\sigma)}\psi_{n}$. 
The result of the dressing procedure in (\ref{2.6}) applied 
to  $\L\in \Xi$ may be parametrized as 
$e^{{\mathrm{ad}} F}(\L)=\partial_x + \Lambda +
\sum_{0<m\neq pn} h_{m}(x)\Lambda^{-m}$.
Then one finds 
\begin{equation}
\H_{m}=\int dx\, h_m(x)=-{1\over m}(-1)^{n} {\mathrm{Tr}}(L^{m \over n}),
\label{3.40}\end{equation} 
whereby the identification of the respective modified systems is complete. 

\medskip
\noindent {\em 3.3.2.~The KdV type system.}
In order to identify the KdV type system 
associated with the above quadruplet as the corresponding 
nonstandard Gelfand-Dickey system, 
we now parametrize the elements of a convenient 
DS gauge $Q_{V}$ in (\ref{2.28}) as 
\begin{equation}
q_{V}(x)= \sum_{k=1}^{n-1}{v_{1} \over n-1}e_{k,k}- v_{1}e_{n,n}- 
\sum_{k=2}^{n-1}(-1)^{k+1}v_{k}e_{n-1,n-k}.
\end{equation}
{}From the linear problem ${\cal L}\Psi =0$,  
we obtain the  eigenvalue equation on $\psi_{n}$  
\begin{equation}
((\partial +{v_{1} \over n-1})^{n-1}+ \sum_{k=2}^{n-1}v_{k}(\partial +
{v_{1} \over n-1})^{n-1-k})(\partial +v_{1})\psi_{n} =
(-1)^n \lambda \psi_{n}
\end{equation}
or equivalently
\begin{equation}
L\tilde{\psi}_{n}=(\partial^{n-1}+\sum_{k=1}^{n-1}u_{k}\partial^{n-1-k})
\partial\tilde{\psi}_{n} = (-1)^n\lambda \tilde{\psi}_{n},\qquad
\tilde{\psi}_{n}= e^{(D^{-1}v_1)}\psi_{n}.
\label{eigen4}
\end{equation}
By the same argument as in paragraph 3.2.2, we conclude that the 
nonlinear bracket on $Q/{\cal N} = Q_{V}$, 
parametrized by the functions $u_k$, is the same as 
the bracket obtained from (\ref{bracb}).
This establishes the desired identification.

\section{Discussion}
\setcounter{equation}{0}

The Gelfand-Dickey system (\ref{1.1}) and its variants in 
(\ref{1.2}), (\ref{3.1}) are reductions of the KP hierarchy
and of its nonstandard counterpart. 
They represent three distinct generalized KdV hierarchies 
whose flows are hamiltonian with respect to the same 
nonlinear Poisson bracket structure given by the
${\cal W}_n\oplus U(1)$ classical extended conformal algebra.
Drinfeld and Sokolov \cite{DS} showed how to view the Gelfand-Dickey
hierarchy in an affine Lie algebraic setting.
Their  derivation used the grade one 
element of the principal Heisenberg subalgebra
of the loop algebra  $sl(n)\otimes\C[\lambda,\lambda^{-1}]$
in a hamiltonian reduction procedure in which 
the interplay between the homogeneous and the principal gradations played 
an important role. 
In this paper, we found that the nonstandard Gelfand-Dickey hierarchy 
of type B (\ref{3.1}) admits a similar derivation in which the homogeneous
gradation is replaced by another gradation. 
In this way, the system in (\ref{3.1}) is interpreted as a special
case of the systems obtained 
by the generalized Drinfeld-Sokolov reduction procedure defined in 
\cite{Prin1}, which associates a KdV type system with a quadruplet 
$(\A, \Lambda, d_1, d_0)$, where the gradations $d_i$ of the loop algebra 
$\A$ satisfy that every element with positive $d_0$-grade is 
positive in the $d_1$-gradation too.

More interestingly, we found that the 
Gelfand-Dickey system of type A (\ref{1.2}) cannot be 
obtained in the framework of \cite{Prin1}.
In fact, we gave a matrix Lax formulation for the 
hierarchy (\ref{1.2}) by using two gradations 
$d_0$ and $d_1$ which satisfy weaker conditions than those above. 
In particular, there may exist elements of the loop algebra $\A$ with 
positive $d_0$ grade and
zero $d_1$ grade, provided they are not orthogonal to the kernel of 
the adjoint action of $\Lambda\in\A$. 
The general Lie algebraic setting which 
we used could be applied  to other cases as well.
A series of such applications was mentioned in Remark 4.

The kernel of the regular semisimple element $\Lambda$ in (\ref{nslambda})
is a maximal abelian subalgebra of the loop 
algebra $\A=sl(n)\otimes\C[\lambda,\lambda^{-1}]$, 
which would become a Heisenberg algebra after introducing 
the central charge. The inequivalent Heisenberg subalgebras
 are classified by the conjugacy classes of the Weyl group of 
$sl(n)$ \cite{kacpeterson}, which are in one-to-one correspondence with 
the partitions of $n$. 
The Heisenberg subalgebra to which $\Lambda$ 
belongs corresponds in the Kac-Peterson classification
 to the conjugacy class in the Weyl group associated with the 
partition $(n-1,1)$. 
Indeed, after a suitable rescaling  of the loop parameter, $\Lambda$ 
may be shown to be equivalent  to
\begin{equation}
\tilde \Lambda=\sum_{i=1}^{n-2}e_{i, i+1}+\tilde \lambda e_{n-1, 1}
\label{conjuglambda}\end{equation}
which is a generator of the principal Heisenberg algebra of the subalgebra 
$sl(n-1)\otimes \C[\tilde \lambda,\tilde \lambda^{-1}]$.
Beside the constrained version (\ref{1.2}) 
of the  nonstandard KP hierarchy 
derived in section 3.2, there exists also a 
constrained version of the standard KP hierarchy
associated with the same Heisenberg subalgebra of $\A$.
This constrained KP hierarchy \cite{cheng}  has a scalar Lax operator 
of the form 
\begin{equation}
\tilde L=\partial^{n-1}+\tilde u_1\partial^{n-2}+\cdots+\tilde u_{n-1}
+\tilde \varphi
(\partial + \tilde w)^{-1}\tilde \chi 
\quad \hbox{with}\quad 
\tilde w=-\tilde u_1,
\label{conskp}\end{equation}
whose derivation in the Drinfeld-Sokolov framework  
is described in \cite{FM}. 
The derivation uses the element $\tilde \Lambda \in \A$ and 
the homogeneous gradation together with a gradation $\tilde d_1$  
for which the condition $\A^0\subset \A_0$ is satisfied. 
More precisely, in \cite{FM}  the $gl(n)$ case is
considered for which $\tilde w$ in (\ref{conskp}) is independent 
of $\tilde u_1$;
neither $\tilde u_1$ nor $\tilde w$ has nontrivial dynamics.
By setting $\tilde w=0$ and conjugating by $\tilde \varphi^{-1}$,  
which is a singular map at the zeros of $\tilde \varphi$, 
one may convert the flows and the Poisson brackets 
of the system based on $\tilde L$ in (\ref{conskp})  into those 
of the system  in (\ref{1.2}) (see \cite{OS}).
Note also that the map $p_A: \tilde L \rightarrow \pa^{-1} \tilde L$ 
connects  the second Poison bracket (but not the flows) of the system in
(\ref{conskp}) to that of a system of the type mentioned in Remark 4.
It should be possible to interpret these  connections 
between the standard and  nonstandard constrained KP hierarchies
as consequences of their closely related affine Lie algebraic origin.

Finally, we wish to stress that in our opinion 
the most interesting problem arising from this paper is to find new input 
data whereby  the general  construction  
of section 2 might give rise to new integrable hierarchies. 
As another problem which remains to be settled,
let us also remark that although the nonstandard Gelfand-Dickey hierarchies 
of type A are known from the scalar Lax formalism to be 
bi-hamiltonian \cite{OS}, we have not yet been able  to identify the first 
Poisson bracket in the matrix Lax formalism.

\bigskip
\bigskip
\bigskip
\noindent 
{\bf Acknowledgments.}
L.F.\ wishes to thank the Laboratoire de Physique 
Th\'eorique, ENS Lyon and the Dublin Institute for Advanced Studies
for hospitality, and also acknowledges support by 
the Hungarian National Science Fund (OTKA) under T016251.

\end{document}